\documentclass[%
 aip,
 amsmath,amssymb,
 reprint,%
]{revtex4-1}
\usepackage{lineno}
\usepackage{graphicx}
\usepackage{dcolumn}
\usepackage{bm}
\usepackage[]{lineno}

\usepackage[utf8]{inputenc}
\usepackage[T1]{fontenc}
\usepackage{mathptmx}
\usepackage{etoolbox}
\usepackage{xcolor}
\usepackage{color,soul}
\makeatletter
\def\@email#1#2{%
 \endgroup
 \patchcmd{\titleblock@produce}
  {\frontmatter@RRAPformat}
  {\frontmatter@RRAPformat{\produce@RRAP{*#1\href{mailto:#2}{#2}}}\frontmatter@RRAPformat}
  {}{}
}%
\makeatother
\begin{document}
\preprint{AIP/123-QED}

\title[Kerr nonlinearity and parametric amplification with an Al-InAs 
superconductor-semiconductor Josephson junction]{Kerr nonlinearity and parametric amplification with an Al-InAs 
superconductor-semiconductor Josephson junction}
\author{Z. Hao}
\author{T. Shaw}
\affiliation{Chandra Department of Electrical and Computer Engineering, University of Texas at Austin, 2501 Speedway, Austin, TX 78712, USA}
\author{M. Hatefipour}
\author{W. M. Strickland}
\author{B. H. Elfeky}
\author{D. Langone}
\author{J. Shabani}
\affiliation{Department of Physics, New York University, 726 Broadway, New York City, NY 10003, USA}
\author{S. Shankar}
\affiliation{Chandra Department of Electrical and Computer Engineering, University of Texas at Austin, 2501 Speedway, Austin, TX 78712, USA}
 \email{shyam.shankar@utexas.edu}

\date{\today}

\begin{abstract}
Nearly quantum limited Josephson parametric amplifiers (JPAs) are essential components in superconducting quantum circuits. However, higher order nonlinearities of the Josephson cosine potential are known to cause gain compression, therefore limiting scalability. In an effort to reduce the fourth order, or Kerr nonlinearity, we realize a parametric amplifier with an Al-InAs superconductor-semiconductor hybrid Josephson junction (JJ). We extract the Kerr nonlinearity of the Al-InAs JJ from two different devices and show that it is three orders of magnitude lower  compared to an Al-$\text{AlO}_\text{X}$ junction with identical Josephson inductance. We then demonstrate a four-wave-mixing (4WM) parametric amplifier made with an Al-InAs junction that achieves more than 20 dB of gain and -119 dBm of compression power, that outperforms single resonant JPAs based on Al junctions. 
\end{abstract}

\maketitle

In superconducting quantum computing, ultra-low-noise Josephson parametric amplifiers (JPAs) are crucial components for high fidelity single-shot readout of superconducting qubits. The most common type of JPA exploits the nonlinearity of the cosine potential energy function of superconductor-insulator-superconductor Josephson junctions (SIS-JJs), to achieve parametric amplification via three-wave or four-wave mixing~\cite{Roy2016,aumentado2020}. An ongoing research theme has been to improve the compression power of these amplifiers in order to process the readout signals arising from many qubits in a large-scale quantum processor. These efforts have typically focused on optimizing the circuit design to maintain only the desired mixing process and minimize parasitic nonlinearities, at the expense of increased design and fabrication complexity~\cite{macklin2015near,Frattini2018,Planat2019understanding,sivak2019,sivak2020,Winkel2020nondegenerate,Planat2020photonic,white2023,Kaufman2023}. An alternative path has been to exploit high kinetic inductance materials ~\cite{Malnou2021-KIPA,parker2022} for amplification, where the intrinsically smaller nonlinearity compared with SIS-JJs results in high compression power. However a trade-off introduced by using kinetic inductance as the source of nonlinearity  is that such devices require extremely high power pumps for amplification which can also limit scalability. Thus, an open question remains as to what is the best source of intrinsic nonlinearity in superconducting quantum circuits which can be exploited to realize a high compression power parametric amplifier while also maintaining other desirable characteristics. 

Recently, Al-InAs superconductor-semiconductor heterostructures are a material system that have received rising attention as a platform for realizing superconducting quantum circuits. Superconductor-semiconductor-superconductor (S-Sm-S) JJs have been demonstrated with this system and used as the basis for the gatemon qubit\cite{casparis2018-gatemon,Yuan2021-gatemon}, voltage-tunable resonators\cite{Strickland2023} and parametric amplifiers~\cite{sarkar2022,butseraen2022,Phan2023,splitthoff2024}. Here, we explore the prospects for this material platform to go beyond traditional SIS-type Al-$\text{AlO}_\text{X}$ JJs to realize ultra-low-noise amplifiers with state-of-the-art performance. We first investigate the Kerr nonlinearity of devices realized with Al-InAs JJs and find that it can be three orders of magnitude lower than identical designs with Al-$\text{AlO}_\text{X}$ junctions. Next, we demonstrate a four-wave-mixing (4WM) parametric amplifier made with an Al-InAs junction that achieves greater than 20 dB of gain, 6 MHz of bandwidth and -119 dBm of compression power.

\begin{figure}
\includegraphics{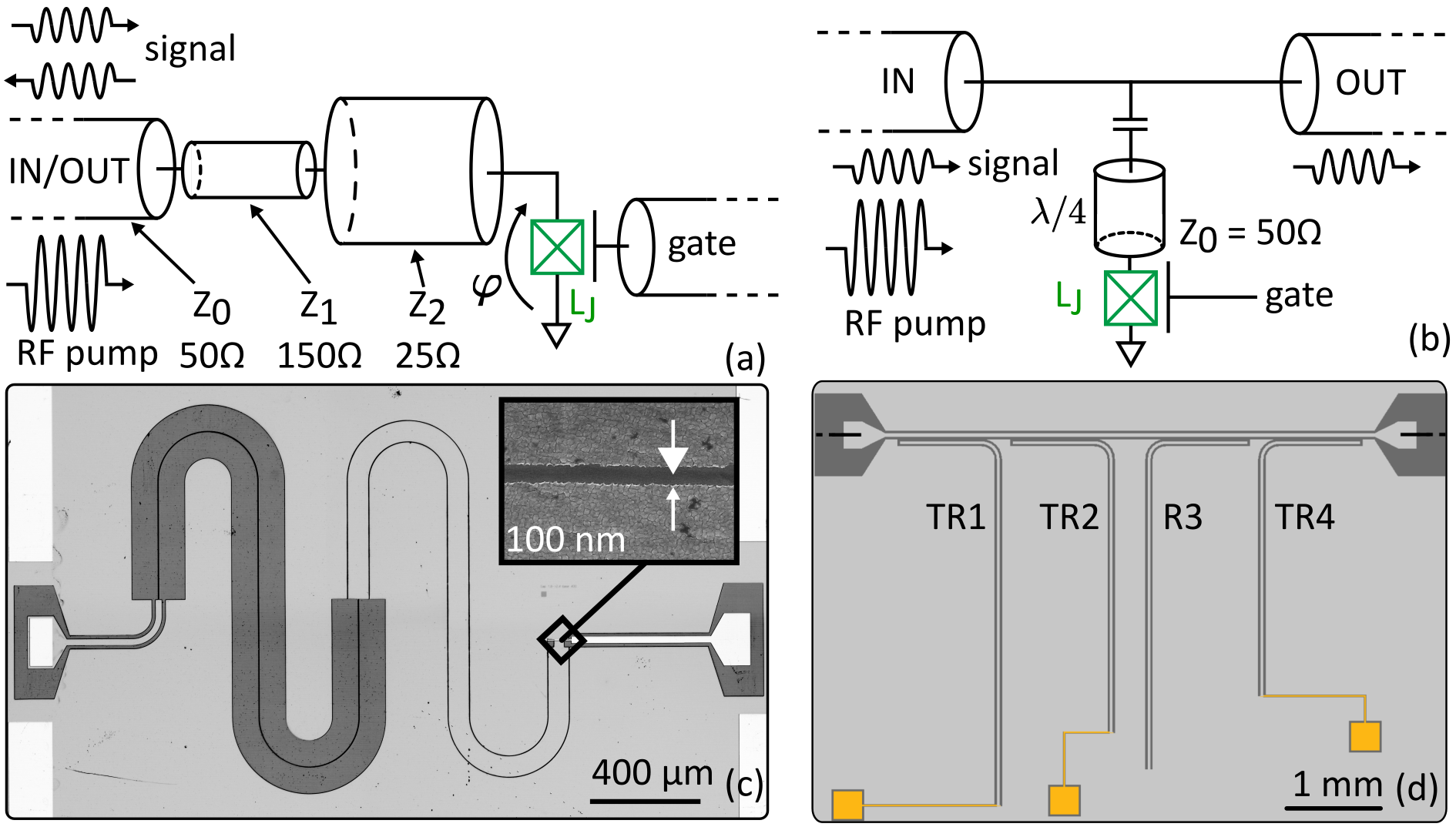}
\caption{\label{fig:figure1}  \textbf{Device design.} (a) Circuit diagram of the direct coupled JJ-FET parametric amplifier. A resonator is formed by two segments of co-planar waveguide transmission lines with characteristic impedance $Z_1 = 150$~$\Omega$, and $Z_2 = 25$~$\Omega$ connected to the Al-InAs Josephson junction (green, $L_J$). (b) Circuit diagram of hanger-style gate-voltage-tunable resonators. (c) Optical image of device A, dark grey: InP substrate, light grey: epitaxially grown Al, white: e-beam evaporated Al.  Inset shows an SEM image of the junction before oxide and gate deposition. (d) Optical image of gate-tunable resonators, where InP is in dark grey, epitaxial Al in light grey, and gate electrodes in gold. Reprinted figure with permission from W. M. Strickland et al, “Superconducting Resonators with Voltage-Controlled Frequency and Nonlinearity.” Physical Review Applied 19, 034021, 2023, https://doi.org/10.1103/PhysRevApplied.19.034021. Copyright 2023 by the
American Physical Society.".
}
\end{figure}

\textit{Devices:} We have measured two devices with different designs shown in figure. \ref{fig:figure1}. Both samples were fabricated on InP wafers on which we realize, by molecular beam-epitaxy, an InAs quantum well in epitaxial ohmic contact with superconducting Al\cite{Shabani2016}. In this Al/InAs heterostructure, an S-Sm-S JJ can be lithographically defined, where the super-current between Al electrodes is carried by electrons in the InAs. The carrier density can be further controlled by a voltage applied on a metal gate electrode, similar to the traditional semiconductor Field Effect Transistor (FET). Therefore, such devices are called Josephson Junction Field Effect Transistor (JJ-FET)\cite{mayer2019superconducting} or Josephson Field Effect Transistor (JoFET)\cite{Phan2023} and we use the former in this report. 


Device A was designed to demonstrate parametric amplification, and is shown in Fig.~\ref{fig:figure1}(a,c). It consists of a JJ-FET directly coupled to the input port via a pair of $\sim \lambda/4$ co-planar waveguide (CPW) transmission lines that act as an impedance matching network. Device B, shown in  Fig.~\ref{fig:figure1}(b,d) and reported in a previous study~\cite{Strickland2023}, consists of a CPW transmission line bus coupled to four hanger-style $\lambda /4$ resonators. Three of the resonators can be controlled by gate voltage to tune their resonance frequencies. In this work, we only focus on the resonator labelled TR2. For both devices we focus on the cases where gate voltages are set to be 0 V. Details on the fabrication of these devices can be found in supplementary material Sec.~4. Table~\ref{tab:table1} lists the measured values for the fundamental resonance frequency $f_r$, and linewidth $\kappa/2\pi$, as well as the values extracted from numerical simulation for the Josephson inductance $L_J$, inductive participation ratio\cite{Manucharyan2007} $p$ and charging energy $E_C$.




\begin{figure}
\includegraphics{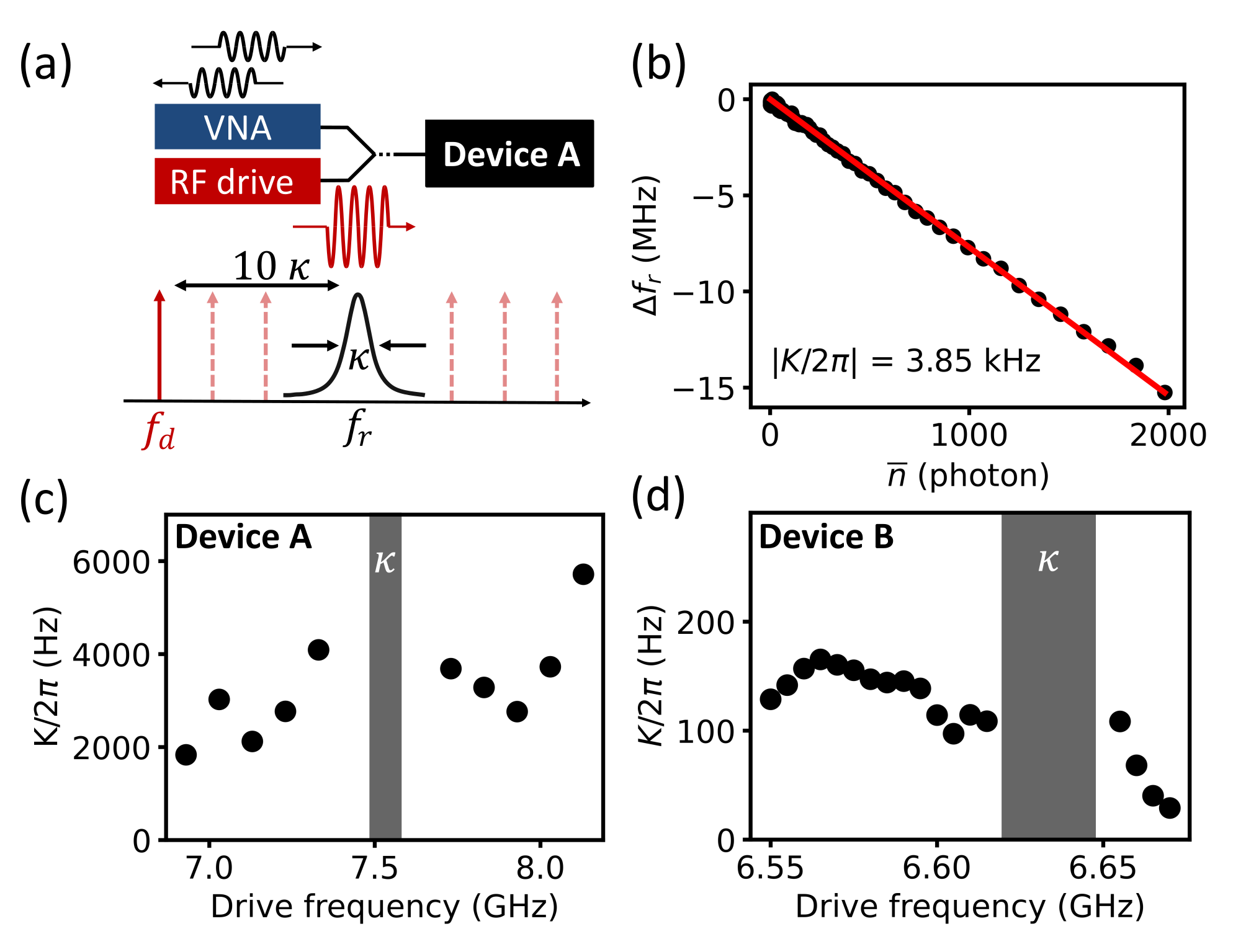}
\caption{\label{fig:figure2}  \textbf{Kerr nonlinearity measurement}. (a) Simplified experiment setup for measurement of Device A. A vector network analyzer (VNA) probes the reflection coefficient of the device, while a signal generator applies a radio-frequency (RF) drive to populate the resonator with photons. Device A was measured in reflection while device B was measured in transmission. The experiment was performed with a sweep of drive frequencies across a range of $\sim 20\kappa$ around the undriven resonance ($\sim 5 \kappa$ for device B).  (b) Extracted resonant frequency shift versus photon number (black dots) for Device A with the drive at $ 7.76$ GHz, along with linear fit (red line) to $\Delta f_r = 2 K \bar{n}$ where $K$ is the Kerr nonlinearity. Measured Kerr nonlinearity of Device A (c) and Device B (d).}
\end{figure}


\textit{Kerr nonlinearity theory and measurements:} An expression for the Kerr nonlinearity of the devices can be found using quantum circuit theory if the current-phase relation (CPR) of the JJ-FET is known. However, unlike the well-known sinusoidal CPR of SIS-type JJs, the CPR for planar super-semi junctions does not, in general, have a closed-form expression. The CPR for similar JJs was measured in Ref.~\cite{Ciaccia2023} and fit to the expression $I(\varphi)=\frac{I_{\mathrm{0}}}{A_N} \frac{\sin (\varphi)}{\sqrt{1-\tau^* \sin ^2(\varphi / 2)}}$ where $I_0$ is the critical current, $\tau^\star$ represents an effective transmission per channel, and $A_N$ is a normalization factor. However, this CPR expression is derived for a single-channel, short-junction JJ\cite{Koops1996}, and its applicability to  our multi-channel, potentially long-junction, JJ is unclear.

Instead, in analogy with the quantum circuit theory of devices with SIS JJs\cite{krantz2019quantum}, we treated the JJ-FET as a generalized nonlinear element whose potential energy is related to the superconducting phase difference between the electrodes $\varphi$ as a Taylor expansion, $U(\varphi) = E_J(c_2\varphi^2/2!-c_4\varphi^4/4! + ...)$. $E_J$ is the Josephson energy, related to the critical current by $E_J = \varphi_0 I_0$ and the Josephson inductance $L_J$ by $L_J = \varphi_0^2/E_J$. $\varphi_0 = \hbar/2e$ is the reduced magnetic flux quantum. When coupled to transmission line structures, the Hamiltonian of the fundamental mode of the circuit is identical to that of a Duffing oscillator, $H/\hbar = \omega_r a^\dagger a + K a^{\dagger 2}a^2$, where $\omega_r = 2\pi f_r$ is the resonance frequency and $K$ is the Kerr nonlinearity. As shown in previous work~\cite{Frattini2018}, desirable properties of a parametric amplifier such as high compression power and bandwidth depend on careful optimization of the Kerr nonlinearity and resonator linewidth. Thus we first characterize the Kerr-nonlinearity for each device.

The Kerr nonlinearity for both devices is extracted from measuring the drive-dependent resonance frequency shift, also known as Stark-shift, via the setup shown in Fig.~\ref{fig:figure2}(a). A drive tone is applied near the resonance and, due to the nonlinearity of the Al-InAs JJ-FET, the resonance frequency lowers as the drive power is increased. Fitting the shift of resonance frequency ($\Delta f$) to the intra-cavity photon number ($\bar{n}$) gives a nearly linear curve at low photon numbers. The slope reveals the Kerr nonlinearity, as shown in Fig.~\ref{fig:figure2}(b). Details on the photon number calibration are given in supplementary material Sec.~2. While typical Stark-shift measurements of nonlinear resonators are done with a single drive frequency a few linewidths away from the undriven resonance, here we chose to sweep the drive frequencies (Fig.~\ref{fig:figure2}(a) lower panel) across a range of $\sim 20 \kappa$ ($\sim 5 \kappa$ for device B) for a more accurate estimate of the error on the extracted $K$.


Figure~\ref{fig:figure2}(c,d) presents the measured Kerr for both devices. The first observation is that there are fluctuations in the measured Kerr nonlinearity versus drive frequency as well as a region approximately $\kappa$ wide near the resonance where no data is shown. We attribute the fluctuations of Kerr extracted away from resonance to variation in cable losses and impedance versus frequency which cause errors in curve fitting and photon number calibration. For drive frequencies within $\sim\kappa$ of the resonance, we observe bifurcation of the Duffing-like nonlinear resonator\cite{Manucharyan2007} as the input power reaches the bifurcation threshold. At these frequencies, we find that the drive is too close to the resonance for the extracted Kerr to be reliable. Together, these indicate that many factors influence the accuracy of the Kerr measurement and suggest that caution should be taken in extracting Kerr values from measurements at a single drive frequency.
On device B, in addition to the Stark-shift measurement, we also performed an inter-modulation distortion (IMD) experiment \cite{sivak2019} to support the Kerr nonlinearity results. In an IMD experiment, two RF tones at the frequencies of 6.6295 and 6.6305 GHz with the same power are applied at the signal input while a spectrum analyzer is used to measure the output power of first and third order intermodulation components. The Kerr nonlinearity can be evaluated from input third-order intercept point (IIP3) as $K = \frac 1 8 \hbar\omega_0 \kappa ^2/\text{IIP3}$ \cite{sivak2019}. We obtain Kerr of 150 Hz at the resonant frequency, which aligns with the Stark-shift measurement on device B. 

\begin{table*}
\caption{\label{tab:table1} Device parameters of two samples, where resonant frequency $f_r$, coupling rate $\kappa$ and Kerr-nonlinearity $K$ are measured while participation ratio $p$, junction inductance $L_J$, charging energy $E_C$ and $c_4/c_2$ ratio are evaluated with a combination of measurements and numerical simulation.}
\begin{ruledtabular}
\begin{tabular}{cccccccc}
 &$f_r$ (GHz)&$\kappa/2\pi$ (MHz)&$p$&
 $L_J$ (nH)&$E_C/\hbar$ (MHz)&$K/2\pi$ (Hz) &$c_4/c_2$\\
\hline
Device A& 7.52& 41.3& 0.29&0.205& 25.6& $3300(\pm1000)$ &$5.3 (\pm 1.6)\times 10^{-3}$\\
Device B& 6.63& 32.3& 0.11&0.200& 51.7& $ 130 
 (\pm 40)$&$1.9(\pm 0.6)\times 10^{-3}$\\
\end{tabular}
\end{ruledtabular}
\end{table*}

Another observation which stands out is that the two devices show approximately an order of magnitude difference of Kerr, even though the junction type and fabrication is nearly identical.To understand this difference, we analyzed factors in the circuit designs that may alter the Kerr coefficient. Table \ref{tab:table1} compares the parameters of the two devices. We extracted $f_r$ and the linewidth $\kappa$ by fitting the linear reflection coefficient. By electromagnetic simulations, described in supplementary material Sec.~1, we extract a lumped element black-box quantization (BBQ)  model\cite{Manucharyan2007, Nigg2012, Frattini2018} to find the equivalent $L$ and $C$ for the fundamental microwave mode coupled to the  junction. The BBQ model gives us the participation ratio of junction $p = L_J/(L+L_J)$, the charging energy $E_C = e^2/2C$ and predicts the Kerr nonlinearity of the mode as $K = (c_4/c_2)p^3E_C/\hbar$, where $c_i = \frac {1} {E_J}\frac {d^nU}{d\varphi^n}|_{\varphi_{min}}$ are the Taylor expansion coefficients of the potential energy function $U(\varphi)$ of the JJ-FET.

The BBQ analysis indicates that both samples have $c_4/c_2$ ratios within a factor of 3 even though the difference of $K$ is more than a factor of 25. While $K$ varies due to the embedding resonator design which set $p$ and $E_C$, the extracted $c_4/c_2$ reveal solely the junction properties. The factor of 3 difference in $c_4/c_2$ ratio between the two devices may be attributed to differences in growth condition and geometry of the junctions. Both devices have a junction length of $\sim 100$ nm; however, junction width of device A and B are $25$~$\mu \text{m}$ and $35$~$\mu \text{m}$ respectively. We can also compare this ratio to that of a SIS Al-$\text{AlO}_\text{X}$ junction with a cosinusoidal potential energy function $U(\varphi) = -\varphi_0^2/L_J\mathrm{cos}(\varphi)$. For an Al-$\text{AlO}_\text{X}$ junction $c_4/c_2$ will be simply 1. The Al-InAs junction, however, has nearly 3 orders of magnitude lower $c_4 / c_2$ ratio, indicating a substantially different and more harmonic potential energy function, or alternatively a more linear current-phase relation, compared to single Al-$\text{AlO}_\text{X}$ junction. While, further study is needed to investigate the relation between junction geometry and higher-order nonlinearities, this substantially lower $c_4/c_2$ ratio is advantageous for demonstrating parametric amplification and other mixing devices as we shown next.

\begin{figure}
\includegraphics{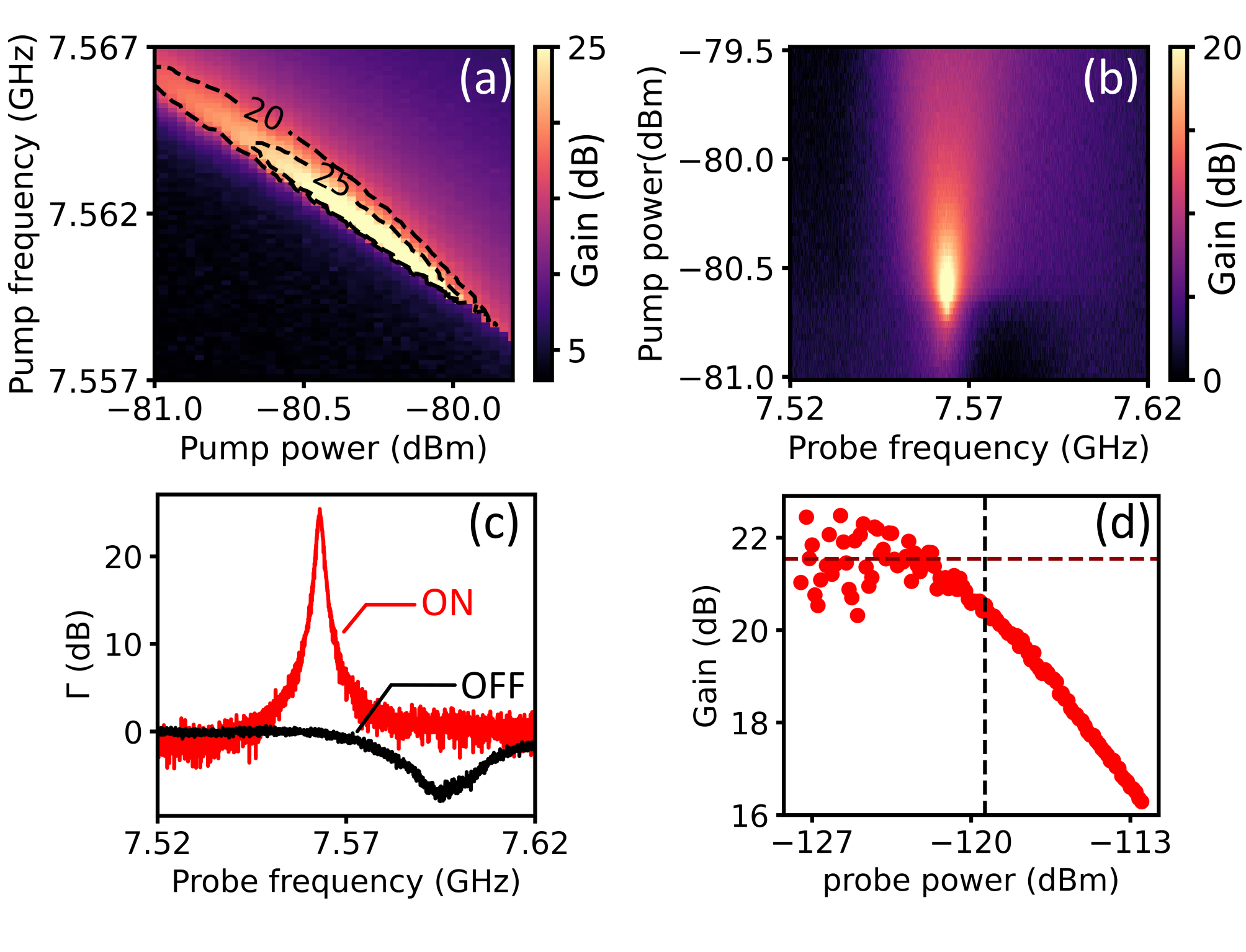}
\caption{\label{fig:figure3}  \textbf{Parametric gain.} (a) Parametric gain as a function of pump power and frequency. Dotted contours  correspond to 20 and 25 dB of gain. (b) Gain versus probe frequency and pump power at fixed pump frequency of 7.562 GHz.(c) Reflection coefficient of the parametric amplifier with the pump applied(red), compared to the un-pumped response (black). The dip in the un-pumped $|\Gamma|$ corresponds to internal $Q\approx660$; the device internal loss could be attributed to dielectric loss in the semiconductor and oxide, and inductive loss in the thin Al film. The peak in gain is at the pump frequency.  (d) Gain as a function of probe power for a representative pump condition of $f_{pump} = 7.562$ GHz, $p_{pump} = -80$ dBm. Dotted line represents compression power $P_{1dB}$ extracted at the pump power where gain drops to 20.6~dB.}
\end{figure}
We demonstrate in Fig.~\ref{fig:figure3}, the observation of parametric amplification using device A. We operate the device in 4WM mode with an RF pump ($f_{pump}$) close to the resonance frequency $f_r$. As theory predicts\cite{Roy2016}, the parametric gain is a function of pump frequency and power. Fig.~\ref{fig:figure3}(a) shows the gain over a sweep of pump conditions. At each pump condition, the reflection coefficient is recorded and the gain is determined by taking the difference in reflection coefficient $\Gamma$ with the pump tone ON versus OFF.  In Fig.~\ref{fig:figure3}(b), $f_{pump}$ is fixed at 7.562~GHz, and the color plot shows the gain is maximized at a certain pump power and probe frequency equal to $f_{pump}$. In Fig.~\ref{fig:figure3}(c), the pump power is also fixed (at -80~dBm) showing that the maximum gain is at $f_{pump}$, which is slightly lower than $f_r$ due to Stark shift, and the 3-dB dynamic bandwidth is about 6~MHz. We also verified that the amplifier noise is near the quantum-limit by measuring the noise visibility ratio (NVR, not shown). The NVR, which is a proxy for the amplifier noise temperature, compares the noise power at the output of our measurement chain with the pump ON versus OFF\cite{bergeal2010phase}. When operated to produce 25~dB gain, the NVR is $\approx 12$~dB above the noise power background, similar to the NVR measured from other nearly-quantum-limited amplifiers in our dilution fridge. These characteristics indicate that the device is behaving as a nearly-quantum-limited, 4WM parametric amplifier, similar to traditional resonant JPAs built with SIS JJs~\cite{aumentado2020}.

A particularly important characteristic of parametric amplifiers is the 1-dB compression power $P_{1dB}$, which characterizes the maximum input probe power below which the amplifier behaves as a linear device. Optimizing the amplifier design for higher compression power is an ongoing research theme in order to scale up superconducting qubit processors~\cite{Frattini2018,sivak2019,sivak2020,white2023,Kaufman2023}. In Fig.~\ref{fig:figure3}(d) we show the measurement of $P_{1dB}$ for device A. We chose a pump condition where the gain saturates at 21.6 dB at very low probe power, then gradually increased probe power until the gain drops by 1 dB. We record this power as $P_{1dB} = -119$~dBm, which is comparable with the compression power of other amplifiers using S-Sm-S JJs\cite{Phan2023,butseraen2022,sarkar2022}.  Moreover, our analysis, described in supplementary material Sec.~3 indicates that an identically designed device with an SIS tunnel junction would have about 10~dB lower compression power since $c_4/c_2 = 1$ for an SIS junction would result in a larger Kerr nonlinearity. While it is possible make JPAs with SIS junctions that have similar Kerr nonlinearity and thus $P_{1dB}$ to device A, this typically requires a more complex design such as using higher critical current junctions or arrays of junctions as well as careful control of participation ratio to get a working amplifier. Moreover, as done with previous resonant JPAs, the compression power of Al-InAs based parametric amplifiers can be further improved by operating the device as a double-pumped 4WM~\cite{kamal2009_signal} or as a three-wave-mixing parametric amplifier, combined with systematic optimization of the participation ratio and bandwidth of the device~\cite{Frattini2018}, or by applying $2\omega_r$ pump at the gate. The intrinsically weaker nonlinearity of Al-InAs JJs compared to Al-$\text{AlO}_\text{X}$ JJs suggests that it is easier to engineer higher compression powers with this material system with correspondingly good prospects for future scaling.



In conclusion, we have investigated the Kerr nonlinearity of the Al-InAs junction and measured that the fourth-order nonlinearity of the potential energy function is three orders of magnitude smaller than comparable Al-$\text{AlO}_\text{X}$ junctions. In the future,  the potential energy function of an Al-InAs JJ to higher orders can be investigated using intermodulation spectroscopy \cite{hutter2010}. Next, a parametric amplifier based on Al-InAs JJ has been built and characterized to be nearly-quantum-limited with more than 20 dB of gain, 6 MHz of bandwidth and compression power of about -119 dBm. Future possibilities including measuring qubits with Al-InAs parametric amplifiers as well as developing three-wave-mixing parametric amplifiers with these junctions.

\section*{Supplementary Material}
See supplementary material for details on derivation of the BBQ models, photon number calibration, microwave simulation for 4WM parametric amplifiers and device fabrication.
\section*{\label{sec:level1}Acknowledgement}
We thank Vladimir Sivak and Nicholas Frattini for  perceptive discussions. This work was supported by the Air Force Office of Scientific Research under grant number FA9550-21-1-0048 and FA9550-20-1-0177. W. Strickland acknowledges the support by the Army Research Office (ARO) and the Laboratory for Physical Sciences (LPS) through the QuaCR Graduate Fellowship (reference number W911NF2110303). This work was performed in part at the University of Texas Microelectronics Research Center, a member of the National Nanotechnology Coordinated Infrastructure (NNCI), which is supported by the National Science Foundation (grant ECCS-2025227).
\section*{DATA AVAILABILITY}
The data that support the findings of this study are available through Zenodo at DOI: 10.5281/zenodo.11075119.

\nocite{*}
\bibliography{manuscript}

\end{document}


\preprint{AIP/123-QED}

\title[]{Supplementary information: Kerr nonlinearity and parametric amplification with an Al-InAs 
superconductor-semiconductor Josephson junction}

\author{Z. Hao}
\author{T. Shaw}
\affiliation{Chandra Department of Electrical and Computer Engineering, University of Texas at Austin, 2501 Speedway, Austin, TX 78712, USA}
\author{M. Hatefipour}
\author{W. M. Strickland}
\author{B. H. Elfeky}
\author{D. Langone}
\author{J. Shabani}
\affiliation{Department of Physics, New York University, 726 Broadway, New York City, NY 10003, USA}
\author{S. Shankar}
\affiliation{Chandra Department of Electrical and Computer Engineering, University of Texas at Austin, 2501 Speedway, Austin, TX 78712, USA}
 \email{shyam.shankar@utexas.edu}

\date{\today}
\maketitle

\renewcommand{\figurename}{Fig.}
\renewcommand{\thefigure}{S\arabic{figure}}
\renewcommand{\thetable}{S\arabic{table}}
\renewcommand{\theequation}{S\arabic{equation}}
\section{Black-box quantization}\label{sec:supp_bbq}
In order to numerically evaluate the Kerr nonlinearity from the physical layout of the devices, we performed electromagnetic simulation with AWR Microwave Office and derived lumped elements models with black-box quantization (BBQ)\cite{Manucharyan2007,Nigg2012} to calculate the  junction participation ratio $p$, and capacitance energy $E_C$. Subsequently the analysis  in Frattini et al.~\cite{Frattini2018} was followed to calculate the Kerr nonlinearity.

\begin{figure}
    \centering
    \includegraphics{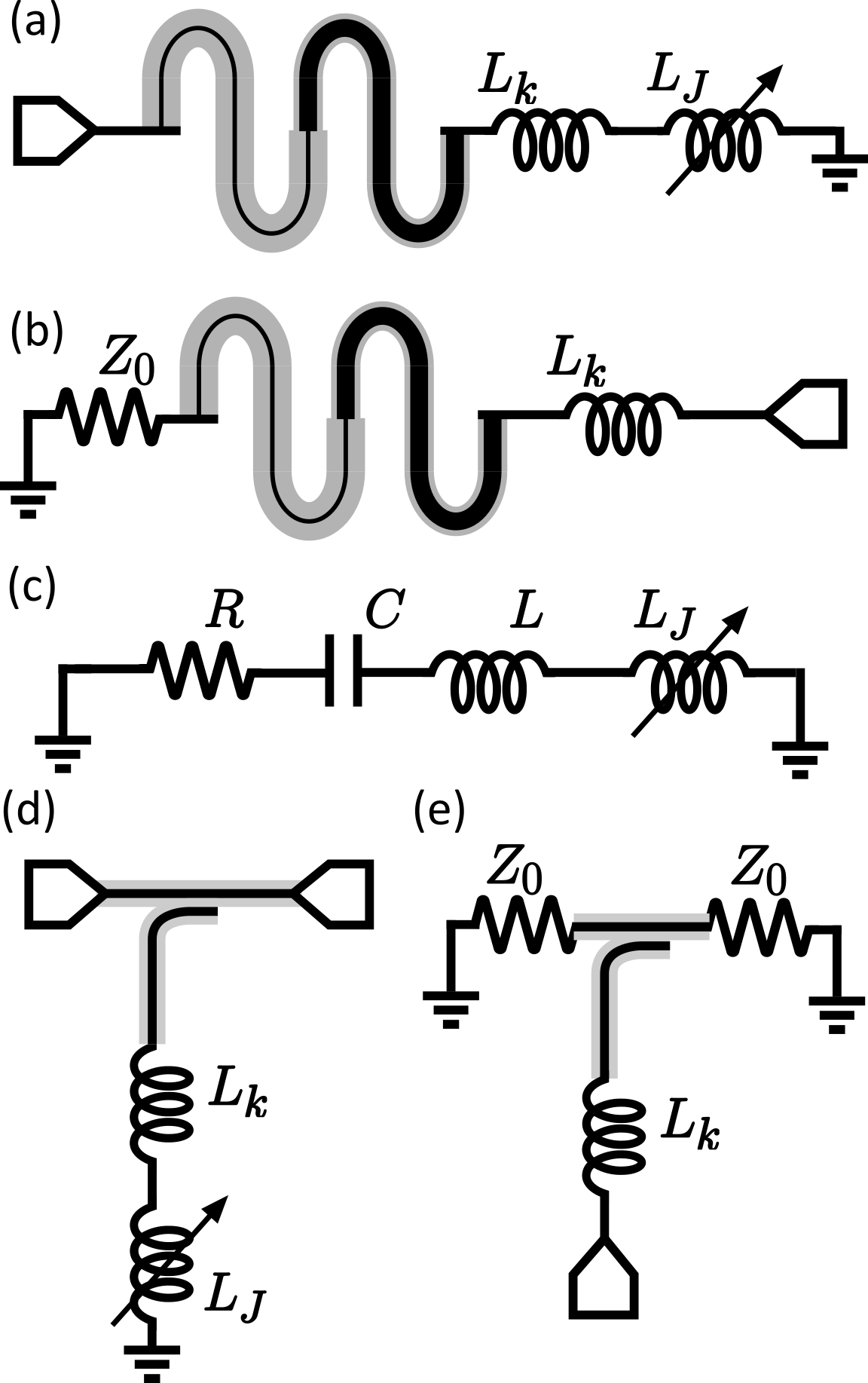}
    \caption{Simplified schematics for device A (a) and B (d) where the Josephson junctions is represented as a tunable linear inductor $L_J$ and the  kinetic inductance is represented as a lumped linear inductor $L_k$.  CPW transmission lines with various $Z_0$ are depicted as gray (dielectric)/black (metal) strips; Ground planes are not shown. (b, e)  Schematics to develop the series BBQ models via EM simulation for device A and B, respectively. (c) A lumped element RLC model that represents a single mode resonance in series with the Josephson junction.}
    \label{fig:sfig1}
\end{figure}

To begin, since the Josephson inductance of the Al-InAs JJ in device A is unknown, we obtained it by matching an electromagnetic simulation of the linear resonance frequency with the experiment results. As shown in Fig.~\ref{fig:sfig1}(a) ,the actual chip layout is used in the simulation, with the JJ replaced by a linear inductor element with value $L_J$. Kinetic inductance  from the $\sim 10$~nm Al film, was modelled as a single series lumped element inductance $L_K$. The value of $L_k$ is evaluated using the physical length ($l$) and width ($w$) of the 25$\Omega$ CPW transmission line as $L_k = \frac l w L_{k,\square}$, where the kinetic inductance per square $L_{k,\square} = 1.012 \text{pH}/\square$ is used as reported in Strickland et al.~\cite{Strickland2023} for a similar aluminum thickness. The reflection coefficient is simulated and the value of $L_J$ is adjusted till the simulated resonance frequency matched that in the experiment. For device B,  $L_J$ was set as reported in previous work\cite{Strickland2023}.

Next, to obtain the BBQ lumped element model parameters, the port position is moved to that shown in Fig.~\ref{fig:sfig1}(b), in order to simulate the the impedance in series with the InAs junction. The resulting impedance function is modelled as a series RLC circuit with $Z(\omega) = (R + j\omega L + 1/(j\omega C))$. The frequency $\omega_0$ at which the imaginary part of the impedance $\text{Im}Z(\omega_0) = \omega_0 L - 1/ \omega_0 C $ becomes zero is given by $\omega_0 = 1/\sqrt{L C}$.  To calculate $L$, we take the derivative of the impedance at $\omega_0$, i.e.\ 
$\text{Im}(Z'(\omega_0)) = L + 1/(\omega_0^2 C) = 2L$. $C$ is calculated as $C = 1/\omega_0^2 L$. Finally, $R$ can be obtained as $R = \text{Re}Z(\omega_0)$, though it is not used in further analysis.

\begin{table}
\caption{\label{tab:stable1}BBQ model parameters for device A and B}
\begin{ruledtabular}
\begin{tabular}{lcccc}
 &  L (nH)&  C (pF)& R($\Omega$)&p\\
 Device A&  0.416&  0.755&  1.174&0.289\\
 Device B&  1.623&  0.474&  0.541&0.110\\
\end{tabular}
\end{ruledtabular}
\end{table}
The resulting effective lumped element circuit model is shown in Fig.~\ref{fig:sfig1}(c). The same approach is adopted to also derive the lumped element model for device B, where the circuit model for the resonance frequency simulation is shown in Fig.~\ref{fig:sfig1}(d) and for the impedance simulation is illustrated in Fig.~\ref{fig:sfig1}(e). The BBQ model parameters for both devices are shown in the Table~\ref{tab:stable1}. Following Frattini et al.~\cite{Frattini2018}, the participation ratio of the junction is $p = L_J/(L_J + L)$ and the charging energy is $E_C = e_c^2/2C$, which are given in the main text.
\section{Photon number calibration}
Obtaining an accurate estimate of the relation between the power set on the microwave signal generator $P_{g}$ to the intra-cavity photon number $\bar n$ is crucial for evaluation of Kerr. We calibrate the input line attenuation by a measurement at room temperature. We estimate that the actual attenuation during Kerr measurement may be systematically smaller by $< 3$~dB which would cause a $<$ factor of 2 systematic error in the extracted Kerr values. 

For device A, a single port reflection mode device, the analysis in Clerk et al.\cite{Clerk2010} gives the photon number as
\begin{equation}
    \bar n = \dfrac{\kappa_{ext}}{\Delta^2 + (\kappa/2)^2}\dfrac{P_g}{\hbar \omega_d A}
\end{equation}
where $\Delta = \omega_d - \omega_r$ is the angular frequency detuning of the drive tone from resonance, $\kappa, \kappa_{ext}$ are the total and external coupling rate and $A$ is the total line attenuation.

For device B, the $\lambda /4$ CPW resonator is coupled to a CPW transmission line, and can be modeled as single-mode `hanger'-style resonance. Following the derivation in Ganjam et al.\cite{ganjam2023surpassing}, and assuming that the coupling is symmetric, the photon number in the resonator is given as
\begin{equation}
    \bar n = \frac{1}{2}\dfrac{\kappa_{ext}}{\Delta^2 + (\kappa/2)^2}\dfrac{P_g}{\hbar \omega_d A}
\end{equation}
where $\Delta$, $\kappa$, $\kappa_{ext}$, $A$ are defined as above.  $\kappa, \kappa_{ext}$ are extracted from fits of the measured linear reflection/transmission coefficient~\cite{mcrae2020_materials}.


\section{Microwave simulation for 4WM parametric amplifier}
In this section, we compare the power-handling performance of SIS and SSmS junctions using numerical harmonic balance simulations. The ratio of the second to fourth Taylor coefficients of the junction potential  $c_4/c_2$ was extracted from Kerr measurements described in the main text. Here we show via numerical simulation that the extracted $c_4/c_2$ ratio for SSmS junctions results in about an order of magnitude higher compression power compared with an identically designed amplifier with an SIS junction.

The harmonic balance simulation was performed with the Cadence AWR APLAC simulator. The junction is modeled current-dependent nonlinear inductance (NLINDA) with $L(I) = L_0 + L_1I + L_2I^2 + L_3I^3 +...$. For an SIS junction, the Taylor expansion coefficients can be calculated directly from the analytic expression $L(I) = L_J/\sqrt{I_c^2-I^2}$ where $I_c = \varphi_0/L_J$. For an SSmS junction, we write 
\begin{equation}
U(\varphi) = E_J \left(\dfrac{c_2}{2!}\varphi^2 - \dfrac{c_4}{4!}\varphi^4\right).
\end{equation}
This gives 
\begin{equation}
\label{eq-i-phi}
I(\varphi)  = \dfrac 1 {\varphi_0} \dfrac {\partial U} {\partial \varphi} = \dfrac {E_J} {\varphi_0} \left( c_2 \varphi - \dfrac{c_4}{3!}\varphi^3\right)
\end{equation}and
\begin{equation}(\varphi) = \left(\dfrac 1 {\varphi_0^2} \dfrac {\partial^2 U} {\partial \varphi^2} \right)^{-1} =  \dfrac {\varphi_0^2}{E_J(c_2 - c_4\varphi^2/2)} =  \dfrac {L_J}{1 - \frac {c_4}{2 c_2}\varphi^2}.
\end{equation}
where we have used $L_J = c_2 E_J /\varphi_0 ^2$.  $L(I)$ can be then derived from a numerical polynomial fit. The values for the $L(I)$ expression for the SSmS junction were calculated using the $c_4/c_2$ ratio extracted from the Kerr measurements.

\begin{figure}
    \centering
    \includegraphics{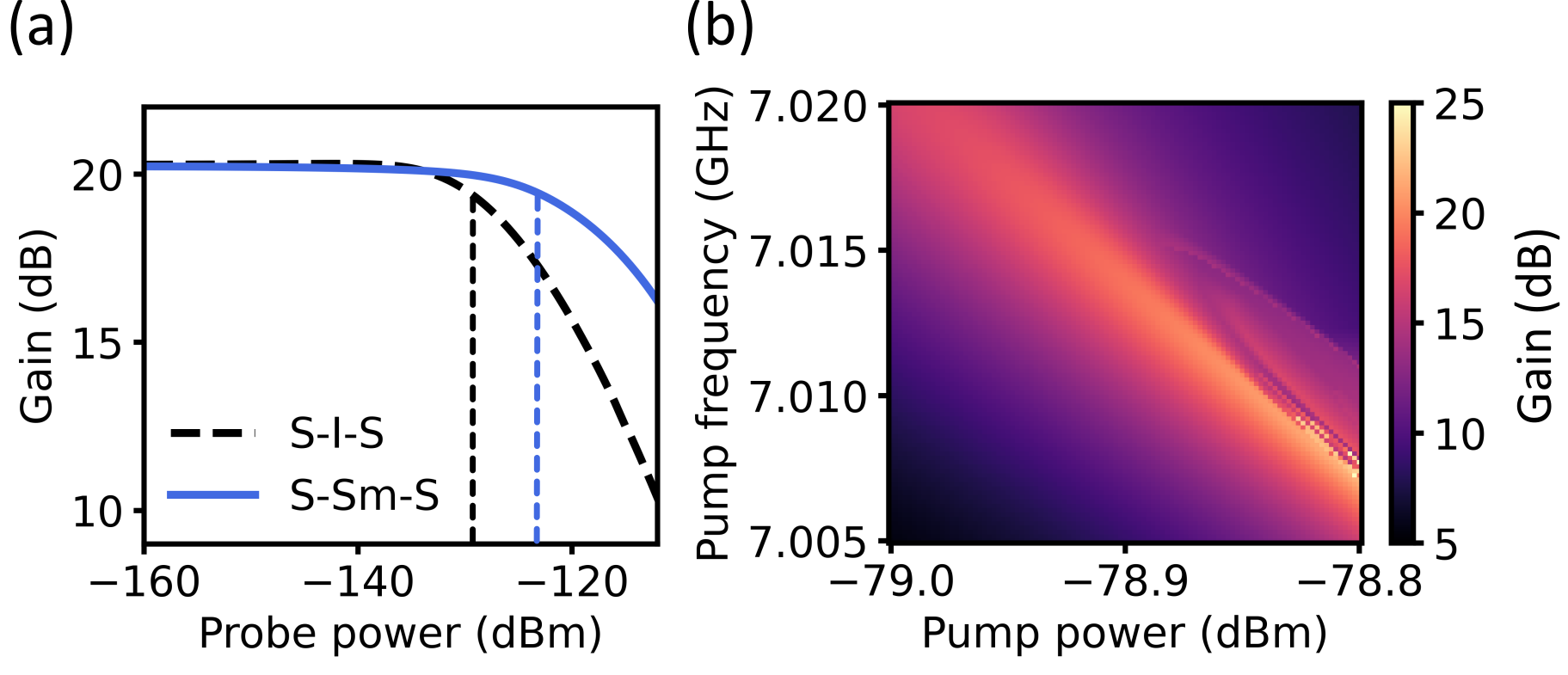}
    \caption{(a) Compression power comparison of parametric amplifiers with SIS (black dashed) and SSmS junction (blue solid) from EM simulation. The gains at very low power are set to about 20 dB. (b)Pump power and frequency sweep to find 20 dB gain for SSmS 4WM amplifier.}
    \label{fig:sfig3}
\end{figure}

Next, nonlinear inductor models (NLINDA) of SIS and SSmS junctions are embedded in the circuits of identical design to simulate 4WM parametric amplifiers. The circuits are the same as device A, they consist of two segments of CPW transmission lines with $150$~$\Omega$ and $25$~$\Omega$ characteristic impedance that couples the junction to the port, with the only difference being the type of junction, expressed via the $L(I)$ expansion. First, pump power and frequency sweeps were performed to search for 20 dB of gain on both circuits. Then we selected the pump condition at 20 dB gain and conduct the probe power sweep to find the 1dB compression points, as shown in Fig.~\ref{fig:sfig3}. Comparing the $P_\text{1dB}$ of SIS and SSmS junction-based JPAs, suggests that the lower nonlinearity of Al-InAs junctions are a good option for achieving high dynamic range in parametric amplifiers.





\section{Device fabrication, packaging and measurement}\label{sec:supp_fab}
In this section, we describe the fabrication details specifically on device A, while noting that device B was fabricated by a similar approach, described further in Strickland et al.~\cite{Strickland2023}. The InAs/InGaAs quantum well and aluminum contact layer is epitaxially grown on an InP wafer~\cite{Shabani2016}. The wafer is then diced into 3~mmx3~mm chips for fabrication. All patterns are imprinted by e-beam lithography (EBL) using a Raith eLine plus with PMMA A3/A4 e-beam resists used for fine/coarse features. CPW center traces and ground plane mesas are defined by a wet etch that removes the Al and all III-V layers down to the InP substrate. The Al is etched by Transene Al Etch Type D under room temperature for 4$\sim$4:30 minutes. III-V semiconductors are etched by a mixture of $\textrm{H}_2 \textrm{O}$ : Citric Acid (1 Mol) : $ \textrm{H}_2\textrm{O}_2 (30\%) : \textrm{H}_3\textrm{PO}_4 (85\%) = 220:55:3:1.3 $ for 15 minutes. Both etch depths are measured using atomic force microscopy (Brucker ScanAsyst), as well as a resistance measurement to ensure that there was no short between disconnected mesas. The JJ is defined by a trench in the Al designed to be 20 nm in the EBL pattern. The trench width after etch is measured under scanning electron microscopy (Zeiss Neon 40) to be around 80 $\sim$ 100 nm. With this trench size and a width of $25 \mu m$, the critical current of the junction is about $2.5 \mu A$. The Al$_2$O$_3$ gate oxide is grown by atomic layer deposition (Fiji F200), with 200 cycles of a single TMA pulse followed by three water pulses. Resulting oxide thickness of $\sim$ 23 nm is  measured by ellipsometry. The gate electrode is $\sim$ 160 nm of Al deposited using e-beam deposition. The deposition is done at an angle of 30$^{\circ}$ to ensure that the Al climbs up the side wall of III-V mesa. An in-situ ion beam etch is performed prior to the metal deposition to clean any residual resist. Finally the gate metal is lifted-off using N-methyl pyrrolidone at 90$^{\circ}$C for 90 minutes, followed by rinse in acetone and isopropyl alcohol. 

Device A was wire bonded to a circuit board (Rogers RO3010) with CPW transmission lines and packaged in a home-built sample holder. Device B is wirebonded and packaged in a commercial QDevil QCage sample holder. The devices are shielded by cryogenic $\mu$-metal shield (Ammuneal A4K) and mounted in a Oxford Triton-500 dilution refrigerator and measured at temperature below 20~mK.

\bibliography{supplementary}